\documentclass{article}%
\usepackage{amssymb}
\usepackage{amsmath}
\usepackage{amsfonts}
\usepackage{graphicx}
\usepackage{indentfirst}%
\begin{document}

\title{Where are ELKO Spinor Fields in Lounesto Spinor Field Classification?}
\author{R. da Rocha$^{1}${\ }and {\ \ }W. A. Rodrigues Jr{\textbf{{\small .}}}$^{2}$\\$^{1}${\footnotesize Gleb Wataghin Physics Institute, Unicamp CP6165}\\{\footnotesize 13083-970 Campinas, SP, Brazil}\\$^{2}${\footnotesize Institute of Mathematics, Statistics and Scientific
Computation}\\{\footnotesize IMECC-Unicamp CP 6065}\\{\footnotesize 13083-859 Campinas, SP, Brazil }\\{\footnotesize Email adresses: \texttt{roldao@ifi.unicamp.br;
walrod@ime.unicamp.br}}}
\maketitle
\date{}

\begin{abstract}
This paper proves that from the \textit{algebraic point of view} ELKO spinor
fields belong together with Majorana spinor fields to a wider class, the
so-called \emph{flagpole} spinor fields, corresponding to the class 5,
according to Lounesto spinor field classification. We show moreover that
algebraic constraints imply that any class 5 spinor field is such that the
2-component spinor fields entering its structure have \textit{opposite}
helicities. The proof of our statement is based on Lounesto \textit{general}
classification of all spinor fields, according to the relations and values
taken by their associated bilinear covariants, and can eventually shed some
new light on the algebraic investigations concerning dark matter.

\end{abstract}

\section{Introduction}

In order to find an adequate mathematical formalism for representing dark
matter, Ahluwalia-Khalilova and Grumiller have recently introduced the
\emph{Eigenspinoren des Ladungskonjugationsoperators} (ELKO) spinor
fields\footnote{Dual-helicity eigenspinors of the charge conjugation
operator.} \cite{allu}, which are shown to belong to a non-standard Wigner
class, and to exhibit non-locality. They claim that an ELKO spinor field is a
new fermion described by a spinor field that has not been identified, in the
Physics literature, with any particle or more general physical entity yet. In
the low-energy limit ELKO comports as a representation of the Lorentz group.
However, mathematicians have already known since sometime ago that all spinors
that are elements of the carrier spaces of the $D^{(1/2,0)}\oplus D^{(0,1/2)}$
or $D^{(1/2,0)}$, or $D^{(0,1/2)}$ representations of $Sl(2,\mathbb{C)}$
belong to one of the six classes found by Lounesto in his theory of the
classification of spinor fields. Such an \textit{algebraic} classification is
based on the values assumed by their bilinear covariants, the Fierz
identities, aggregates and boomerangs (see Eq.(\ref{boom}) below)
\cite{lou1,lou2}. In this paper we prove that from the algebraic point of view
ELKO spinor fields belong to Lounesto class 5 spinor fields, also called
flagpole spinor fields due to the intrinsic flagpole structure they carry. It
is a general property of class 5 spinor fields that they satisfy the Majorana
condition, i.e., if $\psi$ denotes an ELKO spinor field, $\lambda$ a complex
number of unitary modulus and $\mathcal{C}$ the charge conjugation operator
then $\mathcal{C}\psi=\lambda\psi$. In Section 2, after presenting the
bilinear covariants, that completely characterize a spinor field through Fierz
identities and through Fierz aggregates (or boomerangs), Lounesto
classification of spinor fields is reviewed. Section 3 recalls the definition
of ELKO spinor fields \cite{allu} and shows that ELKO is indeed a flagpole
spinor field. We show moreover in Section 4 that algebraic constraints imply
that any class 5 spinor field is such that their 2-component spinor fields
have \textit{opposite} helicities. Now, it is well known that Majorana spinor
fields on Minkowski spacetime are \textit{defined} as eigenvalues of the
charge operator, and it is easy to verify that any spinor field that is
eigenspinor of the charge operator \textit{necessarily} belongs to Lounesto
class 5. So, what differentiates ELKO from Majorana spinor fields? In
\cite{allu} authors quote that the difference is that according to Peskin and
Schroeder \cite{pesksc} and Marshak and Sudarshan \cite{marsud} it is
\textit{imposed} that the 2-component spinor fields of a Majorana spinor field
have the same helicity. Is this reasonable? We briefly discuss such issue.

\section{Bilinear Covariants}

In this paper all spinor fields live in Minkowski spacetime $(M,\eta
,D,\tau_{\eta},\uparrow)$. Here, the manifold $M$ $\simeq\mathbb{R}^{4}$,
$\eta$ denotes a constant metric of signature $(1,3)$, $D$ denotes the
Levi-Civita connection of $\eta$, $M$ is oriented by the 4-volume element
$\tau_{\eta}\in\sec\displaystyle\bigwedge^{4}T^{\ast}M$ and time-oriented by
$\uparrow$. As usual $T^{\ast}M$ denotes the cotangent bundle and $TM$ the
tangent bundle over $M$. By a constant metric we mean the following: let
$\{x^{\mu}\}$ be global coordinates in the Einstein-Lorentz gauge, naturally
adapted to an inertial reference frame \cite{rod1} $\mathbf{e}_{0}%
=\partial/\partial x^{0}$. Let also $\mathbf{e}_{i}=\partial/\partial x^{i}$,
$i=1,2,3$. Then, $\eta(\partial/\partial x^{\mu},\partial/\partial x^{\nu
})=\eta_{\mu\nu}=\mathrm{diag}(1,-1,-1,-1)$. Also, $\{\mathbf{e}_{\mu}\}$ is a
section of the frame bundle $\mathbf{P}_{\mathrm{SO}_{1,3}^{e}}(M)$ and
$\{\mathbf{e}^{\mu}\}$ is its reciprocal frame satisfying $\eta(\mathbf{e}%
^{\mu},\mathbf{e}_{\nu}):=\mathbf{e}^{\mu}\cdot\mathbf{e}_{\nu}=\delta_{\nu
}^{\mu}$. Let $\Xi$ be a section of the principal spin structure bundle
\cite{moro} $\mathbf{P}_{\mathrm{Spin}_{1,3}^{e}}(M)$ such that $s(\Xi
)=\{\mathbf{e}_{\mu}\}$. Classical spinor fields\footnote{Quantum spinor
fields are operator valued distributions, as well known. It is not necessary
to introduce quantum fields in order to know the algebraic classification of
ELKO spinor fields.} carrying a $D^{(1/2,0)}\oplus D^{(0,1/2)}$, or
$D^{(1/2,0)}$, or $D^{(0,1/2)}$ representation of $Sl(2,\mathbb{C)\simeq
}\;\,\mathrm{Spin}_{1,3}^{e}$ are sections of the vector bundle%
\[
\mathbf{P}_{\mathrm{Spin}_{1,3}^{e}}(M)\times_{\rho}\mathbb{C}^{4},
\]
where $\rho$ stands for the $D^{(1/2,0)}\oplus D^{(0,1/2)}$ (or $D^{(1/2,0)}$
or $D^{(0,1/2)}$) representation of $Sl(2,\mathbb{C)\simeq}\;\,\mathrm{Spin}%
_{1,3}^{e}$ in $\mathbb{C}^{4}$. {Other important spinor fields, like Weyl
spinor fields are obtained by imposing some constraints on the sections of
$\mathbf{P}_{\mathrm{Spin}_{1,3}^{e}}(M)\times_{\rho}\mathbb{C}^{4}$. See,
e.g., \cite{lou1,lou2} for details.} Given a spinor field $\psi$ $\in
\sec\mathbf{P}_{\mathrm{Spin}_{1,3}^{e}}(M)\times_{\rho}\mathbb{C}^{4}$ the
bilinear covariants are the following sections of ${\displaystyle\bigwedge
}TM={\displaystyle\bigoplus_{r=0}^{4}}$ ${\displaystyle\bigwedge^{r}%
}TM\hookrightarrow C\mathcal{\ell}(M,\eta)$, where ${\bigwedge}TM$ denotes the
exterior algebra bundle of \textit{multivector} fields and $\mathcal{C\ell
}(M,\eta)$ denotes the Clifford bundle of \textit{multivector }fields of
Minkowski spacetime \cite{moro}:
\begin{align}
\sigma &  =\psi^{\dagger}\gamma_{0}\psi,\quad\mathbf{J}=J_{\mu}\mathbf{e}%
^{\mu}=\psi^{\dagger}\gamma_{0}\gamma_{\mu}\psi\mathbf{e}^{\mu},\quad
\mathbf{S}=S_{\mu\nu}\mathbf{e}^{\mu\nu}=\frac{1}{2}\psi^{\dagger}\gamma
_{0}i\gamma_{\mu\nu}\psi\mathbf{e}^{\mu}\wedge\mathbf{e}^{\nu},\nonumber\\
\mathbf{K} &  =\psi^{\dagger}\gamma_{0}i\gamma_{0123}\gamma_{\mu}%
\psi\mathbf{e}^{\mu},\quad\omega=-\psi^{\dagger}\gamma_{0}\gamma_{0123}%
\psi,\label{fierz}%
\end{align}
with $\sigma,\omega\in\sec%
%TCIMACRO{\dbigwedge ^{0}}%
%BeginExpansion
{\displaystyle\bigwedge^{0}}
%EndExpansion
TM\hookrightarrow C\mathcal{\ell}(M,\eta)$, $\mathbf{J,K}\in\sec%
%TCIMACRO{\dbigwedge ^{1}}%
%BeginExpansion
{\displaystyle\bigwedge^{1}}
%EndExpansion
TM\hookrightarrow C\mathcal{\ell}(M,\eta)$ and $\mathbf{S}\in\sec%
%TCIMACRO{\dbigwedge ^{2}}%
%BeginExpansion
{\displaystyle\bigwedge^{2}}
%EndExpansion
TM\hookrightarrow C\mathcal{\ell}(M,\eta)$. In the formulas appearing in
Eq.(\ref{fierz}) the set $\{\gamma_{\mu}\}$ refers to the Dirac matrices in
chiral representation (see Eq.(\ref{dirac matrices})). Also,%
\[
\{1,\mathbf{e}^{\mu},\mathbf{e}^{\mu}\mathbf{e}^{\nu},\mathbf{e}^{\mu
}\mathbf{e}^{\nu}\mathbf{e}^{\rho},\mathbf{e}^{0}\mathbf{e}^{1}\mathbf{e}%
^{2}\mathbf{e}^{3}\},
\]
\noindent where $\mu,\nu,\rho=0,1,2,3$, and $\mu<\nu<\rho$ is a basis for
$C\mathcal{\ell}(M,\eta)$, and
\[
\{\mathbf{1}_{4},\gamma_{\mu},\gamma_{\mu}\gamma_{\nu},\gamma_{\mu}\gamma
_{\nu}\gamma_{\rho},\gamma_{0}\gamma_{1}\gamma_{2}\gamma_{3}\}
\]
is a basis for $\mathbb{C}(4)$. In addition, these bases satisfy the
respective Clifford algebra relations \cite{lou1}
\begin{align}
\gamma_{\mu}\gamma_{\nu}+\gamma_{\nu}\gamma_{\mu} &  =2\eta_{\mu\nu}%
\mathbf{1}_{4}\text{,}\nonumber\\
\mathbf{e}^{\mu}\mathbf{e}^{\nu}+\mathbf{e}^{\nu}\mathbf{e}^{\mu} &
=2\eta^{\mu\nu},\label{cliffordrelation}%
\end{align}
where $\mathbf{1}_{4}\in$ $\mathbb{C(}4\mathbb{)}$ is the identity matrix, and
$\eta^{\mu\nu}=\mathrm{diag}(1,-1,-1,-1)$. When there is no opportunity for
confusion we shall omit the $\mathbf{1}_{4}$ identity matrix in our formulas.
We observe that the Clifford product of two Clifford fields is denoted by
juxtaposition of symbols. For the orthonormal vector fields $\mathbf{e}^{\mu}$
and $\mathbf{e}^{\nu}$, $\mu\neq\nu,$ their Clifford product $\mathbf{e}^{\mu
}\mathbf{e}^{\nu}$ is equal to the exterior product of those vectors, i.e.,
$\mathbf{e}^{\mu}\mathbf{e}^{\nu}=\mathbf{e}^{\mu}\wedge\mathbf{e}^{\nu
}=\mathbf{e}^{\mu\nu}$.\ \ Also, for $\mu\neq\nu\neq\rho,$ $\mathbf{e}^{\mu}%
{}^{\nu}{}^{\rho}=$\ $\mathbf{e}^{\mu}\mathbf{e}^{\nu}\mathbf{e}^{\rho}$, etc.
More details on our notations, if needed can be found in \cite{moro,rod}.

Since we are interested only in the algebraic classification of spinor fields
it is helpful, in order to consistently perform calculations with algebraic
methods known to the majority of physicists, to introduce operator fields
associated with the bilinear covariant fields. In a fixed spin frame these
operator fields are, for each $x\in M$, mappings $\mathbb{C}^{4}%
\rightarrow\mathbb{C}^{4}.$ They will be represented by the same symbols,
since from this usage (hopefully) no confusion will result. So, in what
follows the bilinear covariants are considered as being the following
\textit{operator} fields:
\begin{align}
\sigma &  =\psi^{\dagger}\gamma_{0}\psi,\quad\mathbf{J}=J_{\mu}\gamma^{\mu
}=\psi^{\dagger}\gamma_{0}\gamma_{\mu}\psi\gamma^{\mu},\quad\mathbf{S}%
=S_{\mu\nu}\gamma^{\mu\nu}=\frac{1}{2}\psi^{\dagger}\gamma_{0}i\gamma_{\mu\nu
}\psi\gamma^{\mu\nu},\nonumber\\
\quad\mathbf{K}  &  =\psi^{\dagger}\gamma_{0}i\gamma_{0123}\gamma_{\mu}%
\psi\gamma^{\mu},\quad\omega=-\psi^{\dagger}\gamma_{0}\gamma_{0123}\psi.
\label{11}%
\end{align}
\noindent In the case of the electron, described by Dirac spinor fields
(classes 1, 2 and 3 below), $\mathbf{J}$ is a future-oriented timelike current
vector which gives the current of probability. This means that the Clifford
product of $\mathbf{J}\in\sec{\displaystyle\bigwedge^{1}}TM\hookrightarrow
C\mathcal{\ell}(M,\eta)$ with itself, i.e., $\mathbf{J}^{2}$ is such that
\begin{equation}
\mathbf{J}^{2}=J_{\mu}\mathbf{e}^{\mu}J_{\nu}\mathbf{e}^{\nu}=J_{\mu}J_{\nu
}\frac{1}{2}(\mathbf{e}^{\mu}\mathbf{e}^{\nu}+\mathbf{e}^{\mu}\mathbf{e}^{\nu
})=\eta^{\mu\nu}J_{\mu}J_{\nu}=J_{\mu}J^{\mu}>0.
\end{equation}
\noindent Of course, if $\mathbf{J}:M\ni x\mapsto\mathbb{C(}4\mathbb{)}$ is
interpreted as a vector (field) operator we have $\mathbf{J}^{2}=J_{\mu}%
J^{\mu}\mathbf{1}_{4}$. In this case writing $\mathbf{J}^{2}>0$ means $J_{\mu
}J^{\mu}>0$.

Moreover, the bivector $\mathbf{S}$ is associated with the distribution of
intrinsic angular momentum, and the spacelike vector $\mathbf{K}$ is
associated with the direction of the electron spin. For a detailed discussion
concerning such entities, their relationships and physical interpretation, and
generalizations, see, e.g., \cite{cra,lou1,lou2,holl,hol}.

The bilinear covariants satisfy the Fierz identities \cite{cra,lou1,
lou2,holl,hol}
\begin{equation}
\mathbf{J}^{2}=\omega^{2}+\sigma^{2},\quad\mathbf{K}^{2}=-\mathbf{J}^{2}%
,\quad\mathbf{J}\cdot\mathbf{K}=0,\quad\mathbf{J}\wedge\mathbf{K}%
=-(\omega+\sigma\gamma_{0123})\mathbf{S}. \label{fi}%
\end{equation}
\noindent and also satisfy\footnote{Note that \textbf{S}$^{-1}$ exists of
course only if $\omega$ and $\sigma$ are \textit{not} simultaneously null.}:
\begin{align}
\mathbf{S}\llcorner\mathbf{J}  &  =\omega\mathbf{K}\qquad\mathbf{S}%
\llcorner\mathbf{K}=\omega\mathbf{J},\qquad(\gamma_{0123}\mathbf{S}%
)\llcorner\mathbf{J}=\sigma\mathbf{K},\nonumber\\
(\gamma_{0123}\mathbf{S})\llcorner\mathbf{K}  &  =\sigma\mathbf{J}%
,\qquad\mathbf{S}\llcorner\mathbf{S}=-\omega^{2}+\sigma^{2},\qquad
(\gamma_{0123}\mathbf{S)}\llcorner\mathbf{S}=-2\omega\sigma,\nonumber\\
\mathbf{J}\mathbf{S}  &  =-(\omega+\sigma\gamma_{0123})\mathbf{K}%
,\quad\mathbf{K}\mathbf{S}=-(\omega+\sigma\gamma_{0123})\mathbf{J}%
,\quad\mathbf{S}\mathbf{J}=(\omega-\sigma\gamma_{0123})\mathbf{K}\nonumber\\
\mathbf{S}\mathbf{K}  &  =(\omega-\sigma\gamma_{0123})\mathbf{J}%
,\qquad\mathbf{S}^{2}=(\omega-\sigma\gamma_{0123})^{2}=\omega^{2}-\sigma
^{2}-2\omega\sigma\gamma_{0123},\nonumber\\
\mathbf{S}^{-1}  &  =-\mathbf{S}\frac{(\sigma-\omega\gamma_{0123})^{2}%
}{(\omega^{2}+\sigma^{2})^{2}}=\frac{\mathbf{K}\mathbf{S}\mathbf{K}}%
{(\sigma^{2}+\omega^{2})^{2}}. \label{FIERZ}%
\end{align}

In the formulas above $\cdot$ denotes the scalar product and $\wedge$ refers
to the exterior product, while $\llcorner$ to the \textit{right} contraction
product of Clifford fields (or Clifford operators). For details, please
consult, e.g., \cite{lou1,lou2,rod1}. Introduce the complex multivector field
$Z\in\sec\mathbb{C}\ell(M,\eta)$ (where $\mathbb{C}\ell(M,\eta)$ denotes the
complexified spacetime Clifford bundle, in which the typical fiber is
$\mathbb{C\otimes R}_{1,3}\simeq\mathbb{R}_{4,1}$ \cite{moro}) and the
corresponding complex multivector\textit{ operator }(represented by the same
letter):
\begin{equation}
Z=\sigma+\mathbf{J}+i\mathbf{S}+i\mathbf{K}\gamma_{0123}+\omega\gamma_{0123.}
\label{boom}%
\end{equation}
When the multivector operators $\sigma,\omega,\mathbf{J},\mathbf{S}%
,\mathbf{K}$ satisfy the Fierz identities, then the complex multivector
operator $Z$ is denominated a \emph{Fierz aggregate}, and, when $\gamma
_{0}Z^{\dagger}\gamma_{0}=Z$, which means that $Z$ is a Dirac self-adjoint
aggregate\footnote{It is equivalent to say that $\omega,\sigma,\mathbf{J}%
,\mathbf{K},\mathbf{S}$ are real multivectors.}, $Z$ is called a
\emph{boomerang}.

A spinor field such that \emph{not both} $\omega$ and $\sigma$ are null is
said to be regular. When $\omega=0=\sigma$, a spinor field is said to be
\textit{singular}. In this case the Fierz identities are in general replaced
by the more general conditions \cite{cra} (which obviously also holds for
$\omega,\sigma\neq0$). These conditions are:
\begin{align}
Z^{2}  &  =4\sigma Z,\qquad Z\gamma_{\mu}Z=4J_{\mu}Z,\qquad Zi\gamma_{\mu\nu
}Z=4S_{\mu\nu}Z,\nonumber\\
Zi\gamma_{0123}\gamma_{\mu}Z  &  =4K_{\mu}Z,\qquad Z\gamma_{0123}Z=-4\omega Z.
\end{align}

Now, any spinor field (regular or singular) can be reconstructed from its
bilinear covariants as follows. Take an arbitrary spinor field $\xi$
satisfying $\xi^{\dagger}\gamma_{0}\psi\neq0.$ Then the spinor field $\psi$
and the multivector field $Z\xi$, differ only by a phase. Indeed, it can be
written as
\begin{equation}
\psi=\frac{1}{4N}e^{-i\alpha}Z\xi, \label{a1}%
\end{equation}
\noindent where $N=\frac{1}{2}\sqrt{\xi^{\dagger}\gamma_{0}Z\xi}$ and
$e^{-i\alpha}=\frac{1}{N}\xi^{\dagger}\gamma_{0}\psi$. For more details see,
e.g., \cite{cra, qtmosna}.

Lounesto spinor field classification is given by the following spinor field
classes \cite{lou1,lou2}, where in the first three classes it is implicit that
$\mathbf{J}$\textbf{, }$\mathbf{K}$\textbf{, }$\mathbf{S}$ $\neq0$:

\begin{enumerate}
\item $\sigma\neq0,\;\;\; \omega\neq0$.

\item $\sigma\neq0,\;\;\; \omega= 0$.

\item $\sigma= 0, \;\;\;\omega\neq0$.

\item $\sigma= 0 = \omega, \;\;\;\mathbf{K}\neq0,\;\;\; \mathbf{S}\neq0$.

\item $\sigma= 0 = \omega, \;\;\;\mathbf{K}= 0, \;\;\;\mathbf{S}\neq0$.

\item $\sigma= 0 = \omega, \;\;\; \mathbf{K}\neq0, \;\;\; \mathbf{S} = 0$.
\end{enumerate}

\noindent The current density $\mathbf{J}$ is always non-zero. Type 1, 2 and 3
spinor fields are denominated \textit{Dirac spinor fields} for spin-1/2
particles and type 4, 5, and 6 are respectively called \textit{flag-dipole},
\textit{flagpole} and \textit{Weyl spinor fields}. Majorana spinor fields are
a particular case of a type 5 spinor field. It is worthwhile to point out a
peculiar feature of types 4, 5 and 6 spinor fields: although $\mathbf{J}$ is
always non-zero, we have $\mathbf{J}^{2}=-\mathbf{K}^{2}=0$. We shall see,
below, that the bilinear covariants related to an ELKO spinor field, satisfy
$\sigma=0=\omega,\;\;\mathbf{K}=0,\;\;\mathbf{S}\neq0$ and $\mathbf{J}^{2}=0$.

Lounesto proved that there are \textit{no} other classes based on distinctions
between bilinear covariants. So, ELKO spinor fields must belong to one of the
six classes.

Before ending this section we remark that the sum of two spinor fields
belonging to a given Lounesto class is not necessarily a spinor field of the
same class, as it is easy to verify.

\section{ELKO Spinor Fields}

In this section we explore in details the algebraic properties of ELKO spinor
fields as defined in \cite{allu}.

A ELKO spinor field $\Psi$ corresponding to a plane wave with momentum
$p=(p^{0},\mathbf{p)}$ can be written, without loss of generality, as
$\Psi=\psi e^{-i{p\cdot x}}$ (or $\Psi=\psi e^{i{p\cdot x}}$) with
\begin{equation}
\psi=\binom{i\Theta\phi_{L}^{\ast}(\mathbf{p})}{\phi_{L}(\mathbf{p})},
\label{1}%
\end{equation}
\noindent where, given the rotation generators denoted by ${\mathfrak{J}}$,
the Wigner's spin-1/2 time reversal operator $\Theta$ satisfies $\Theta
\mathfrak{J}\Theta^{-1}=-\mathfrak{J}^{\ast}$. It is useful to choose
$i\Theta=\sigma_{2}$, as in \cite{allu}, in such a way that it is possible to
express
\begin{equation}
\psi=\binom{\sigma_{2}\phi_{L}^{\ast}(\mathbf{p})}{\phi_{L}(\mathbf{p})}.
\label{01}%
\end{equation}
\noindent Here, as in \cite{allu}, the Weyl representation of $\gamma^{\mu}$
is used, i.e.,
\begin{equation}
\gamma_{0}=\gamma^{0}=%
\begin{pmatrix}
0 & \mathbf{1}_{2}\\
\mathbf{1}_{2} & 0
\end{pmatrix}
,\quad-\gamma_{k}=\gamma^{k}=%
\begin{pmatrix}
0 & -\sigma_{k}\\
\sigma_{k} & 0
\end{pmatrix}
, \label{dirac matrices}%
\end{equation}
\noindent where
\begin{equation}
\sigma_{1}=%
\begin{pmatrix}
0 & 1\\
1 & 0
\end{pmatrix}
,\quad\sigma_{2}=%
\begin{pmatrix}
0 & -i\\
i & 0
\end{pmatrix}
,\quad\sigma_{3}=%
\begin{pmatrix}
1 & 0\\
0 & -1
\end{pmatrix}
\end{equation}
\noindent are the Pauli matrices.

Omitting the subindex of the spinor $\phi_{L}(\mathbf{p})$, which is denoted
heretofore by $\phi$, the left-handed spinor field $\phi_{L}(\mathbf{p})$ can
be represented by
\begin{equation}
\phi=\binom{\alpha(\mathbf{p})}{\beta(\mathbf{p})},\quad\alpha(\mathbf{p}%
),\beta(\mathbf{p})\in\mathbb{C}. \label{0}%
\end{equation}
\noindent\noindent Now using Eqs.(\ref{11}) it is now possible to calculate
explicitly the bilinear covariants for ELKO spinor fields:
\begin{align}
\sigma &  =\psi^{\dagger}\gamma_{0}\psi=0,\\
& \nonumber\\
\omega &  =-\psi^{\dagger}\gamma_{0}\gamma_{0123}\psi=0\label{16a}\\
& \nonumber\\
\mathbf{J}  &  =J_{\mu}\gamma^{\mu}=\psi^{\dagger}\gamma_{0}\gamma_{\mu}%
\psi\gamma^{\mu}\nonumber\\
&  =2(\alpha\beta^{\ast}+\alpha^{\ast}\beta)\gamma^{1}+2i(\alpha^{\ast}%
\beta-\alpha\beta^{\ast})\gamma^{2}+2(\beta\beta^{\ast}-\alpha\alpha^{\ast
})\gamma^{3}\nonumber\\
&  \quad+2(\alpha\alpha^{\ast}+\beta\beta^{\ast})\gamma^{0},\label{16}\\
& \nonumber\\
\mathbf{K}  &  =K_{\mu}\gamma^{\mu}=\psi^{\dagger}i\gamma_{123}\gamma_{\mu
}\psi\gamma^{\mu}=0, \label{17}%
\end{align}
\medbreak
\begin{align}
\mathbf{S}  &  =\frac{1}{2}S_{\mu\nu}\gamma^{\mu\nu}=\frac{1}{2}\psi^{\dagger
}\gamma_{0}i\gamma_{\mu\nu}\psi\gamma^{\mu\nu}\nonumber\\
&  =\frac{i}{2}((\alpha^{\ast})^{2}+(\beta^{\ast})^{2}-\beta^{2}-\alpha
^{2})\gamma^{02}+\frac{1}{2}((\alpha^{\ast})^{2}+(\beta^{\ast})^{2}+\beta
^{2}+\alpha^{2})\gamma^{31}\nonumber\\
&  \quad+\frac{1}{2}((\beta^{\ast})^{2}+\beta^{2}-(\alpha^{\ast})^{2}%
-\alpha^{2})\gamma^{01}+\frac{i}{2}(-\beta^{2}-\alpha^{2}+(\alpha^{\ast}%
)^{2}+(\beta^{\ast})^{2})\gamma^{02}\nonumber\\
&  \quad+(\alpha\beta+\alpha^{\ast}\beta^{\ast})\gamma^{03}+\frac{i}{2}%
(\alpha\beta-\alpha^{\ast}\beta^{\ast})\gamma^{12}+\frac{i}{2}(\beta
^{2}-\alpha^{2}+(\alpha^{\ast})^{2}-({\beta^{\ast})}^{2})\gamma^{23}.
\label{18}%
\end{align}
\noindent From the formulas in Eqs.(\ref{16}, \ref{17}) it is trivially seen
that that
\begin{equation}
\mathbf{J}\cdot\mathbf{K}=0. \label{a2}%
\end{equation}
\noindent Also, from Eq.(\ref{16}) it follows that
\[
\mathbf{J}^{2}=0,
\]
\noindent and it is immediate that all Fierz identities introduced by the
formulas\ in Eqs.(\ref{fi}) are trivially satisfied. It also follows directly
from Eq.(\ref{18}) (or easier yet, using the formula for $\mathbf{S}^{2}$ in
Eq.(\ref{FIERZ}) and Eq.(\ref{16a})) that $\mathbf{S}^{2}=0.$

Now, any flagpole spinor field is an eigenspinor of the charge conjugation
operator \cite{lou1,lou2}, here represented by $\mathcal{C}\psi=-\gamma
^{2}\psi^{\ast}$. We must have:
\begin{equation}
-\gamma^{2}\psi^{\ast}=\lambda\psi,\quad|\lambda|\text{ }=1.
\end{equation}
\noindent where $\lambda$ is a complex number of unitary modulus. \noindent
Using Eq.(\ref{01}) it follows that
\begin{align}
-\gamma^{2}\psi^{\ast}  &  =%
\begin{pmatrix}
0 & \sigma_{2}\\
-\sigma_{2} & 0
\end{pmatrix}
\binom{(\sigma_{2}\phi^{\ast})^{\ast}}{\phi^{\ast}}\nonumber\\
&  =\binom{\sigma_{2}\phi^{\ast}}{-\sigma_{2}\sigma_{2}^{\ast}\phi}\nonumber\\
&  =\psi.
\end{align}
\noindent Now, recall that a Majorana spinor field (in Minkowski spacetime) is
defined as an eigenvector of the charge operator, with $\lambda=\pm1$. It
follows, as can be easy verified that its structure implies immediately that
it must be a flagpole spinor field, i.e., of Lounesto class 5 spinor fields.
\cite{lou1,lou2}. So ELKO spinor field \textit{belongs} to the same class as
Majorana spinor fields, i.e., class 5 spinor fields, by Lounesto spinor field
classification. So, the question arises: what is the difference between ELKO
and Majorana spinor fields?

\section{Helicities}

Consider any class 5 spinor field $\psi=\binom{\phi_{1}}{\phi_{2}}$. We
already stated that any such $\psi$ satisfy the equation $\mathcal{C}%
\psi=\lambda\psi$ with $\lambda\lambda^{\ast}=1$. Such condition implies that
$\lambda\phi_{1}=\sigma_{2}\phi_{2}^{\ast}$.

Let as usual $\mathbf{\sigma\cdot\hat{p}}$ be the helicity operator acting on
2-component spinor fields. Suppose now that $\mathbf{\sigma\cdot\hat{p}}%
\phi_{2}=\phi_{2}$ (respectively, $\mathbf{\sigma\cdot\hat{p}}\phi_{1}%
=\phi_{1}$). Then a trivial calculation shows that\ $\mathbf{\sigma\cdot
\hat{p}}\phi_{1}=-\phi_{1}$ (respectively $\mathbf{\sigma\cdot\hat{p}}\phi
_{2}=-\phi_{2}$), i.e., the 2-component spinor fields presented in the
structure of a class 5 spinor field have necessarily \textit{opposite} helicities.

Of course, this is also the case of an ELKO spinor field, since we have just
proved that they belong to class 5 spinor fields. We are now prepared to give
the answer to the question formulated at the end of the previous section,
according to Ahluwalia-Khalilova and Grumiller \cite{allu}. They asserted that
the difference between ELKO and Majorana spinor fields resides in the fact
that the 2-component spinor fields entering the structure of a Majorana spinor
field have the \textit{same} helicity. They attribute this statement, e.g., to
Peskin and Schroeder \cite{pesksc} and Marshak and Sudarshan \cite{marsud}. Of
course, this assumption if used by \cite{pesksc,marsud} or by any other author
must be considered completely ad hoc from the \textit{algebraic} point of
view. There is no justification for it, except an eventual desire to give to
Majorana particles a well-defined helicity, something that is not endorsed by
the Mathematics of spinor fields. Reading carefully Peskin and Schroeder's
book \cite{pesksc} we found that those authors propose as an
exercise\footnote{Exercise 3.4, page 73 of \cite{pesksc}.} the possibility of
writing a field equation for a 2-component spinor field of \textit{definite
}helicity (with Grassmann algebra-valued entries) \textit{encoding} the
contents of a Majorana field. In part (e) of that exercise those authors\ call
the 2-component spinor field (of definite helicity) a Majorana field. However,
the true Majorana spinor field is a 4-component spinor field, eigenvector of
the charge operator and thus, as already proved, it must be composed by two
2-component spinor fields of opposite helicities. At the heart of the issue it
is a real confusion between the concepts of \textit{chirality} and
\textit{helicity} for massive fermions. Indeed, we can find papers, e.g., one
by Hannestad \cite{hanne} where it is stated that the Majorana
\textit{quantum} spinor field is indeed without definite chirality (i.e.,
\textquotedblleft it is a linear combination of left handed and right handed
parts\textquotedblright, these parts understood as chiral parts of a spinor
field) but that the \textit{quantum state} of a Majorana particle may be of
definite helicity. Hannestad endorses his statement quoting the theory of
Majorana particles as derived in the book by Mohapatra and Pal \cite{mopal}
and also on the book by Kim and Pevsner \cite{kipe}. Also, Plaga \cite{plaga}
stated that Majorana fermions may have states of definite helicity. However,
adding power to the confusion he indeed exhibits a \textquotedblleft Majorana
field\textquotedblright\ with definite helicity, something that according to
our view is equivocated. Plaga also stated that physical states cannot have
definite chirality. A complete discussion of these issues will be postponed to
another paper. Here we only quote that Ahluwalia-Khalilova and Grumiller
\cite{allu} showed that adhering to the correct mathematical result leads to
interesting physical consequences, as, e.g., the issue of non-locality (see
also \cite{22}).

\section{Concluding Remarks}

We showed that ELKO spinor fields belong to the class of flagpole, class 5,
spinor fields, according to Lounesto classification. We showed moreover that
algebraic constraints imply that any class 5 spinor field is such that their
2-component spinor fields have \textit{opposite} helicities. The statements
that \cite{allu} attributes to \cite{pesksc,marsud} asserting that Majorana
spinor fields, (a particular class 5 spinor field) are such that their
2-component spinor fields have the same helicity seems to be ad hoc. Moreover,
it is easy to verify that\footnote{As first observed by Ahluwalia-Khalilova
and Grumiller in \cite{allu}.} while the \emph{anticommutator} between the
charge conjugation and parity operators acting on a Dirac spinor field is
equal to zero, the \emph{commutator} of those operators acting on an ELKO
spinor field (which do not satisfy Dirac equation) is also zero. ELKO dynamics
is to be analyzed in a forthcoming paper. Also, a relation between Lounesto's
classification and Wigner's classification of spinor fields (which plays an
important role in \cite{allu}) needs some further study and will be presented elsewhere.

Finally, we take the opportunity to call the reader's attention to the fact
that no use has been made until now (to the best of our knowledge) of class 4
spinor fields. Eventually they may be the important spinor fields to describe
dark matter and/or dark energy. This possibility will be to explored elsewhere.

\section*{Acknowledgements}

The authors are grateful to Professor D. V. Ahluwalia-Khalilova for having
asked to us to study the nature of ELKO spinor fields and to Dr. R. A. Mosna
for his helpful suggestions. R. da Rocha is also grateful to CAPES for
financial support.

\end{document}